# Identifying Unvaccinated Individuals in Canada: A Predictive Model

Kevin Dick
Systems and Computer Engineering
Carleton University
Ottawa, Ontario
Email: kevin.dick@carleton.ca

Ardyn Nordstrom
Department of Economics
Carleton University
Ottawa, Ontario
Email: ardyn.nordstrom@carleton.ca

*Abstract*—Recently, the media and public health officials have become increasingly aware of the rise in anti-vaccine sentiment. Vaccinations have numerous health benefits for immunized individuals as well as for the general public through herd immunity. Given the rise in immunization-preventable diseases, a consequence of people opting out of their routine vaccinations, we determined that Canadian health data can identify individuals over the age of 60 who chose not to get vaccinated (80.1% negative predictive value) and individuals under the age of 60 who have recently been vaccinated (96.4% positive predictive value). Using the 2009 - 2014 Canadian Community Health Surveys (CCHS), a probit model identified the variables that were most commonly associated with flu vaccination outcomes. Of 1,381 variables, 47 with the most significant marginal effects were selected, including the presence of diseases (*e.g.* diabetes and cancer), behavioural characteristics (*e.g.* smoking and exercise), exposure to the medical system (*e.g.* whether the individual gets a regular check-up), and a person's living situation (*e.g.* having young children in the household). These variables were then used to generate a Random Forest classification model, trained on the 2009-2013 dataset, and tested on the 2014 dataset. We achieved an overall accuracy of 87.8% between the two final models, each using 25 classification trees with bounded depth of 20 nodes, randomly selecting from all 47 variables. With the two proposed policies, this model can be leveraged to efficiently allocate vaccination promotion efforts. Additionally, it can be applied to future surveys, only requiring 3.6% of the variables in the CCHS for successful prediction.

## I. INTRODUCTION

The 2013 Childhood National Immunization Coverage Survey Government found that as many as 23% of Canadian children do not have their basic immunizations (*e.g.* whooping cough) [1]. Less than one third of Canadians receive their flu shot each year [2]. The dramatic decrease in vaccination rates compromises herd immunity, which is established and maintained when a significant portion of the population is vaccinated against a particular pathogen, thereby decreasing the spread of contagious diseases. The poor flu vaccine coverage is particularly dangerous for seniors since the flu can evolve into more serious health conditions. The Canadian government has recognized this as a serious issue and, in the most recent budget, allocated $25 million to vaccination efforts in Canada [3].

In this work, we sought to determine whether large survey datasets could be used to predict whether individuals were at highest risk of not receiving their flu shot. Section II of this paper describes the dataset used in this analysis: the Canadian Community Health Survey (CCHS), for which we have consistent data from 2009 to 2014.

Section III outlines the methodology employed to develop this model. This was a two-step process, which involved first using probit models and statistical tests to determine which of the 1,381 variables were most significant in predicting who got their flu shot. Identifying relevant variables using a probit model was particularly useful since probit models are effective in predicting binary dependent variables[4], [5]. In this model, the dependent variable referred to whether someone received his or her flu shot in the previous 12 months (a binary variable). A probit model based on the cumulative distribution function of the Normal distribution, is preferred over other nonlinear models (such as logit models, which are based on the cumulative distribution function of the Exponential distribution) when there is a large sample size [4], [5]. This research used data sets with over 60,000 records for each sample year thereby satisfying the large-sample condition. We generated and compared several classification models and opted for a random forest classifier, which was optimized. The random forest classification method is a machine learning technique wherein a combination of tree predictors are constructed from randomly, independently, sampled (with replacement) vectors using identical sampling distributions. As the number of trees in our forest grows, the generalization error converges and the impact of noise is reduced [6]. They are widely applicable to all types of data, given that they do not have any linearity requirements for their features, and can be applied to catagorical (binary) features. Through the use of bagging or boosting, random forests can additioanlly incorporate high dimensional feature spaces and very large training sets, making them robust for big data analytics.

The use of sets of covariates for the categorization or prediction of binary outcomes for making binary decisions is broadly applied in the field of economics and in fields of science. Binary outcome predictions generally support decision-making, which may also be binary in nature. The use of binary and continuous variables simultaneously in the development of a prediction model has been demonstrated to be successful for accurate prediction of outcomes [7]. The evaluation of binary and continuous variables for vaccination outcome prediction produces a binary result (whether an individual has recently been vaccinated or not) which in turn serves as an indicator for the subsequent binary decision (whether to promote reception of the flu vaccine or not) specific to that individual.

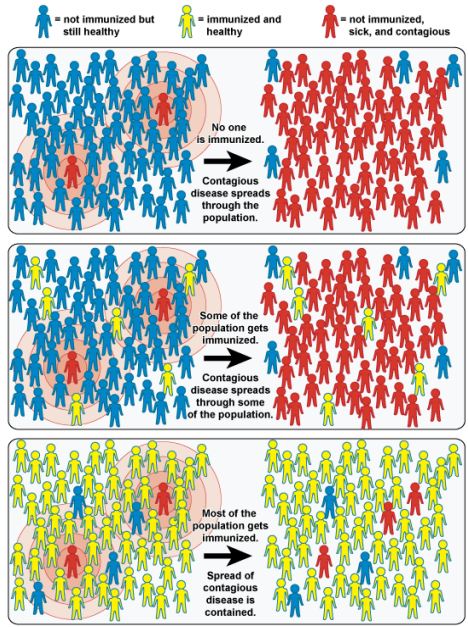

Fig. 1. Herd Immunity. When a sufficeintly large percentage of the population is vaccinated against a particular pathogen, the liklihood of transmission of that infectious agent from a diseased individual to a non-immunized individual is dramatically reduced as a consequence of the preventative nature of the vaccinated surrounding community. Source: [8]

In section IV, we describe the model results and given the age sensitivity of these results, implications of these results are discussed in section V and specifically outlines two policies for how this model could be leveraged by public health officials to direct resources within the population.

## II. Dataset

The Canadian Community Health Surveys (2009-2014) were used for this analysis. This survey is designed to acquire information from approximately 65,000 households in Canada every year and includes approximately 1,381 variables in each dataset[1][9]. This survey collects demographic details about surveyed households, as well as details about their health conditions, access and use of medical facilities, and behavioural traits (such as their driving, smoking, and drinking habits). This data is published by Statistics Canada with detailed documentation.

The variables were structured in distinct groups, and these groups were manually curated to broadly identify those suited to our application. Entire groups of variables were excluded due to insufficient information (*e.g.* mammogram and prostate exam questions were only asked to a small portion of the sample, resulting in insufficient information to include these variables in the analysis). After manually eliminating groups of variables, categorical variables (*e.g.* province, gender, and whether someone received a flu shot in the last twelve months) were converted into binary variables for each category to make interpretation more intuitive. Many of the categorical variables are assigned arbitrary values in the CCHS making the interpretation of them meaningless. For example, the province variable in the CCHS assumed values from 10 - 70 to indicate different provinces but these values are intrinsically meaningless. Making multiple binary variables for each province allowed the unique effects of each province to be evaluated separately.

Some sample restrictions had to be made to exclude observations with missing data. These restrictions included the elimination of individuals who did not provide flu shot information (1,698 observations) as well as observations missing education information (4,285 observations), missing income information (5,960 observations), and missing marital status information (131 observations) footnoteAll these exclusion counts refer to the exclusions associated with the 2012 dataset..

Table I describes some of the main summary statistics associated with the CCHS datasets (2012 and 2014 have been included to demonstrate that the samples are relatively consistent across time). In 2012, 37% of the sample received their flu shot. This rose to 42% in 2014. Moreover, the proportion of individuals who reported used their phone while driving was 0.3% and 0.5% in 2012 and 2014, respectively. The proportion of individuals who reported smoking daily was 21% in 2012, and 20% in 2014. 55% of the observations in these featured samples were female, 54% of observations were married, and approximately 10% had a child aged 0-5 in their household.

Of the health conditions reported in this survey, 8% of individuals in each sample year reported having asthma, and 22% and 26% of individuals reported having arthritis in 2012 and 2014, respectively. Additionally, 8% and 9% of the samples reported having a mood disorder in 2012 and 2014, respectively, while 7% of individuals reported having an anxiety disorder each year.

## III. Methodology

### A. Variable Identification

To identify variables that were important in predicting the likelihood that an individual received their flu shot in the previous twelve months, multiple probit models were estimated. In making this prediction, this model considered demographic information about individuals that are likely associated with flu shot decisions (*e.g.* young children in the household), as well as health information associated with flu shot decisions (*e.g.* having asthma, having a family doctor, getting regular check ups). In addition to these features, this analysis considered behavioural characteristics that may be associated with "risky" behaviour (*e.g.* smoking, driving without a seatbelt) since choosing not to get a flu shot could also be considered a "risky" decision. Therefore, there may be underlying personality characteristics that motivate all these types of behaviour that must be incorporated in the model. By implementing these probit models, this analysis applied Student T-tests and grouped Chi-squared tests to identify which individual variables and groups of variables were significantly associated with the vaccination outcome, which was defined as whether someone received their flu shot in the last twelve months. After entire groups of variables were excluded from the sample due to poor data quality, four iterative probit models were developed, and groups of variables were tested for significance with each iteration until only groups of variables with significant explanatory power remained in the model. Table II describes the variables that were excluded in the preliminary probit models, as well as the T-statistics and Chi-Squared statistics that led to the conclusion[2] that these variables had no significant explanatory power in predicting flu vaccinations. Variables pertaining to an individual's illicit drug use, their proclivity to engaging in risky sexual behaviours, and whether or not they have insurance were all found to be statistically and intrinsically insignificant (once their marginal effects were evaluated).

---

[1]Approximate because certain variables, such as religiosity or suicidal behaviour, were not included in the survey every year

TABLE I
SAMPLE MEANS WITH STANDARD DEVIATIONS IN BRACKETS

| Variable | 2012 | 2014 |
|---|---|---|
| Got Flu Shot in Last Year | 0.366 | 0.418 |
|  | (0.482) | (0.493) |
| *Behaviour* | | |
| Drives Without Seatbelt | 0.003 | 0.005 |
|  | (0.053) | (0.068) |
| Uses Phone While Driving | 0.005 | 0.009 |
|  | (0.067) | (0.095) |
| No Attempt to Find Family Doctor | 0.051 | 0.046 |
|  | (0.220) | (0.210) |
| Has a Family Doctor | 0.867 | 0.874 |
|  | (0.339) | (0.332) |
| Daily Smoker | 0.212 | 0.196 |
|  | (0.409) | (0.397) |
| BMI | 26.571 | 26.821 |
|  | (5.304) | (5.404) |
| Frequent Exercise | 0.687 | 0.682 |
|  | (0.464) | (0.466) |
| Occasional Exercise | 0.145 | 0.142 |
|  | (0.352) | (0.349) |
| *Demographics* | | |
| Female | 0.550 | 0.550 |
|  | (0.498) | (0.497) |
| Immigrant | 0.145 | 0.143 |
|  | (0.352) | (0.350) |
| Married/ Common-law | 0.535 | 0.537 |
|  | (0.499) | (0.499) |
| Divorced/ Separated | 0.234 | 0.249 |
|  | (0.423) | (0.432) |
| Child (age 0-5) | 0.111 | 0.101 |
|  | (0.314) | (0.301) |
| Child (age 6-11) | 0.105 | 0.098 |
|  | (0.306) | (0.298) |
| *Province of Residence* | | |
| British Columbia | 0.122 | 0.123 |
|  | (0.328) | (0.328) |
| Alberta | 0.088 | 0.096 |
|  | (0.284) | (0.295) |
| Saskatchewan | 0.055 | 0.057 |
|  | (0.227) | (0.232) |
| Manitoba | 0.052 | 0.058 |
|  | (0.223) | (0.233) |
| Ontario | 0.332 | 0.318 |
|  | (0.471) | (0.466) |
| Québec | 0.203 | 0.191 |
|  | (0.402) | (0.393) |
| New Brunswick | 0.043 | 0.039 |
|  | (0.202) | (0.194) |
| Nova Scotia | 0.038 | 0.043 |
|  | (0.192) | (0.203) |
| Prince Edward Island | 0.015 | 0.016 |
|  | (0.120) | (0.125) |
| Newfoundland | 0.029 | 0.033 |
|  | (0.168) | (0.179) |
| *Health Conditions* | | |
| Asthmatic | 0.082 | 0.083 |
|  | (0.274) | (0.275) |
| Arthritis | 0.224 | 0.255 |
|  | (0.417) | (0.436) |
| Cancer | 0.037 | 0.028 |
|  | (0.190) | (0.165) |
| Diabetes | 0.085 | 0.098 |
|  | (0.278) | (0.297) |
| Heart Disease | 0.073 | 0.076 |
|  | (0.261) | (0.265) |
| Mood Disorder | 0.083 | 0.090 |
|  | (0.275) | (0.286) |
| Anxiety Disorder | 0.065 | 0.071 |
|  | (0.246) | (0.258) |
| Personal Income | 39,686 | 41,899 |
|  | (28,058) | (28,462) |
| **Observations** | **43,759** | **45,300** |

TABLE II
SIGNIFICANCE TESTS FOR EXCLUDED VARIABLES

| Variable | T-Statistic | Joint Chi-Squared Statistic |
|---|---|---|
| Uses Birth Control | 1.62 | |
| History of Sexually Transmitted Disease | -1.19 | 3.81 |
| Illicit Drug User | -1.57 | - |
| Weekday Drinking | -1.63 | - |
| Repetitive Strain Injury in the Last Year | 1.61 | - |
| Has Dental Insurance | 1.62 | - |

Once these variables had been eliminated, the final probit model had the following form:

$$flushot_i = \alpha + EDUC_i\beta_1 + DRIVE_i\beta_2 + \beta_3 female_i \\ + MARSTAT_i\beta_4 + HEALTH_i\beta_5 \\ + X_i\beta_6 + \varepsilon_i \quad (1)$$

where

- *flushot* refers to whether individual *i* received their flu shot in the last 12 months
- *EDUC* refers to a set of binary variables indicating individual *i*'s highest level of education
- *DRIVE* refers to a set of binary variables indicating whether individual *i* engages in risky driving behaviours
- *female* refers to a binary variable indicating if individual *i* is female
- *MARSTAT* refers to a set of binary variables indicating individual *i*'s marital status

[2]Using the relevant test statistics at the 5% significance level.

- *HEALTH* refers to a set of binary variables indicating individual *i*'s health conditions
- *X* refers to a matrix of individual *i*'s other characteristics (*e.g.* age, income, demographics)

The choice to use a probit model is motivated by the fact that the dependent variable in the regression is a "dummy" variable that indicates whether an individual received their flu shot in the previous year. A probit model is well suited to models with a categorical dependent variable when there are at least several thousands observations [5], [4] as found in [10].

A summary of the marginal effects of the variables in the final model are described in Table III. It should be noted that the exclusion of these variables did not have a substantial effect on the explanatory power of the probit model. The full (initial) probit model was associated with an $R^2$ value of 0.1520. Once the final exclusions had been made, this $R^2$ only fell by 0.0001 to 0.1519. This supported the conclusion that these features did not have a significant relationship with whether or not someone received their flu shot in the past 12 months.

Table III shows that individuals who report driving with a cell phone are 11.27% less likely to recieve their flu shot (all else equal), while individuals who do not wear a seat belt are 8.9% less likely to get their flu shot (all else equal)[3]. Individuals who smoke are 7.25% less likely to recieve their flu shot compared to non-smokers, while individuals who get a regular check-up are 9.37% more likely to recieve their flu shot. Individuals with diseases such as asthma or cancer are 8.89% and 5.57% more likely to recieve their flu shot, respectively, than comparable healthy individuals. Moreover, individuals in Québec are the least likely to recieve their flu shot (14.73% less likely than comparisons in Ontario), while individuals in Nova Scotia are the most likely to their flu shot (5.45% more likely than Ontario comparisons). Individuals who have children under the age of 6 are 11.47% more likely to get their flu shot than individuals without young children, and women are 6.15% more likely than male comparisons to get their flu shot.

### B. Classification Model Development

The variables identified using the probit models became the binary features considered in the development of our classifier. Several experiments were performed to generate and optimize the final classification model, a decision node splitting data by age 60 and two Random Forest classifiers trained on different data sets employing 25 decision trees, each bounded by a decision depth of 20 levels, with random feature selection across all 47 features.

*1) Determination of the Classification Model:* The Weka Machine Learning software [11] was used to manipulate the data sets, generate the classification models, and predict outcomes of our test set (the outcome being whether or not the individual has recently recieved the flu vaccine). The majority of our experiments were performed using the 2011 and 2012 datasets given that they were the most currently available set at the time. On March 16, 2016, the 2013 and 2014 data sets were released and tested against our model.

---

[3]These marginal effects are based on the 2012 survey, however, the results are very similar for other survey years as well.

TABLE III
MARGINAL EFFECTS ON PROBABILITY OF GETTING FLU SHOT

| Top Influencing Variables | Marginal Effect on Probability of Getting Flu Shot |
|---|---|
| *Behaviour* | |
| Uses a cell phone while driving | 11.27% less likely*** |
| Not wearing a seatbelt while driving | 8.9% less likely** |
| Frequently exercises | 2.57% more likely*** |
| Chooses health for food content | 6.56% more likely*** |
| Has strong social relationships | 4.15% more likely*** |
| Regularly smokes | 7.29% less likely*** |
| Getting a regular check-up | 9.37% more likely*** |
| Having a family doctor | 7.35% more likely*** |
| Didn't attempt to find a doctor | 6.59% less likely*** |
| *Demographics* | |
| Having a young child (age 0-5) | 11.47% more likely than people with no children*** |
| Female | 6.18% more likely than men*** |
| Married or common-law | 2.94% less likely than single comparisons*** |
| Divorced/widowed/separated | 4.21% less likely to get vaccinated than single comparisons*** |
| Post-secondary grad. | 2.35% more likely*** |
| *Health Conditions* | |
| Diabetic | 11.89% more likely*** |
| Asthmatic | 8.89% more likely*** |
| Heart disease | 8.52% more likely*** |
| Cancer | 7.15% more likely*** |
| Arthritis | 5.57% more likely*** |
| *Province- Ontario as Baseline* | |
| British Columbia | 12.96% less likely*** |
| Alberta | 9.22% less likely*** |
| Saskatchewan | 11.23% less likely*** |
| Manitoba | 7.67% less likely*** |
| Québec | 14.73% less likely*** |
| New Brunswick | 0.99% less likely* |
| Nova Scotia | 5.45% more likely*** |
| Prince Edward Island | 8.33% less likely*** |
| Newfoundland and Labrador | 8.41% less likely*** |

\* Significant at 10%;
\*\* Significant at 5%;
\*\*\* Significant at 1%

When determining the baseline classification results for different types of classifiers, a variety of methods were considered to identify whether a certain method excelled over another for our data. One classification technique was selected from the following classification families: probabilistic classifiers applying Bayes' theorem, decision trees, ensemble methods, neural network classification, and non-parametric models.

- **Naive Bayes:** Selected for its simplicity and to determine the baseline of classification when naively assuming independence between features
- **J48 Decision Tree:** Employing the C4.5 algorithm, this classifier applies information entropy when determining decision node segregation
- **Random Forest:** An ensemble model of random tree classifiers with built-in bagging ("bootstrap aggregation" to improve accuracy and avoid over-fitting)
- **Multilayer Perceptron:** A feed forward artificial neural network
- **K*:** A non-parametric, instance-based method using entropy as a measure of distance

Each classification model was initially tested with its default Weka parameters and the performance metric of each was compared across years. The most meaningful metrics for our evaluation are the Positive Predictive Value (PPV), given as the ratio of correctly identified postives by our classifier, the Negative Predictive Value (NPV), described as the ratio of correctly identified negatives by our classifier, and accuracy (ACC) which refers to the ratio of correctly classified individuals. A true positive (TP) refers to an individual who has correctly been identified as having recently been vaccinated, a true negative (TN) refers to an individual who has correctly been identified as *not* having recently been vaccinated. A false positive (FP) refers to an individual identified by our model as having been recently vaccinated, when in reality then have not, and a false negative (FN) referes to an individual identified by our model as *not* having been vaccinated when in reality they had.

$$PPV = \frac{TP}{TP+FP} \quad (2)$$

$$NPV = \frac{TN}{TN+FN} \quad (3)$$

$$ACC = \frac{TP+TN}{TP+TN+FP+FN} \quad (4)$$

The model Receiver Operating Characteristic value was also determined and reported, however for the consideration of model optimization, the PPV, NPV and ACC metrics were used. Attribute evaluation algorithms (*e.g.* $\chi^2$ statistical test) were applied to the data in conjunction with a ranking function (i.e. Ranker) to motivate feature selection to determine whether a subset of features could be used to generate a simpler model. A bottom-up approach, wherein our model was compared to previous results with the iterative inclusion of additional features, was adopted. The incremental incorporation of features was performed in accordance to the ranking of variables as evaluated across four considered attribute evaluation methods: Symmetrical Uncertainty, Chi-Squared, Gain Ratio, and Information Gain.

*2) Determination of the Age-Split Boundary:* To further optimize our model, we investigated the nature of the misclassified individuals. The majority of FP and FN were found to originate in "younger" individuals, motivating the development of two classifiers; one optimized for classification of older individuals and a second for younger individuals. The data was segregated at an age-split boundary such that a subset of the data fell on or below that boundary and the remaining became the elder subset. This was accomplished by sorting the data by age and segregating the data at various years. Determination of the operational boundary was accomplished by splitting the data for each year over the range of 18 to 72 and then verifying the performance metrics for each division when training on that subset, and testing on the equivalent subset in the independent test set. A coarse division of age by a decade was considered in thise work, beginning with an age-split at 30 through to 70. We considered the PPV for the younger population given that we preferentially wish to identify those individuals in the younger population that have previously received their flu vaccination. We considered the NPV for the older population given that we preferentially wish to identify those individuals in senior populations that have not previously received their flu vaccination and are therefore at greater risk of adverse health effects should they become infected. The two classifiers were then optimized independently for their respective metric of interest. Optimization was performed by adjusting three parameters: number of decision trees in the random forest, the maximum depth permitted of each decision tree in the random forest, and the number of features over which random feature selection could occur.

*3) Development of the Final Model:* The final model incorporates all data from 2009-2013 for training and the independent 2014 dataset for testing and two random forest classifiers. The random forest classifier for individuals over the age of 60 was trained using only individuals in the 2009-2013 data set that were above the age of 60, while the random forest classifier for individuals under the age of 60 was trained using the entire data sets. The deployment of our model applies an initial decision node to segregate individuals by age 60 and their classification is then determined by that respective random forest classifier.

TABLE IV
COMPARISON OF PERFORMANCE METRICS OVER FIVE CLASSIFIERS USING NINE SELECTED FEATURES. [4]

|          | Naive Bayes | J48 Decision Tree | Random Forest | Multilayer Perceptron | K*    |
|----------|-------------|-------------------|---------------|-----------------------|-------|
| PPV      | 0.784       | 0.861             | 0.851         | 0.862                 | 0.874 |
| NPV      | 0.548       | 0.484             | 0.523         | 0.471                 | 0.464 |
| Accuracy | 0.698       | 0.723             | 0.721         | 0.719                 | 0.724 |
| ROC      | 0.729       | 0.710             | 0.764         | 0.795                 | 0.752 |

## IV. RESULTS

The baseline performance metrics when considering the original 2011 (training) and 2012 (testing) data sets with a subset of the highest ranking nine of the 47 features identified are given in Table IV and the comparable results using all 47 features are reported in Table V. Each classification method performed similarly over each dataset and improved consistently when incorporating additional features. The PPV was found to be consistently higher than the NPV, which did not generally improve above random (50%). The random forest classification method was selected since it is a robust (*i.e.* avoids over-fitting the training set) and stable classification technique that is configurable to optimize and implement. The top ranking features from the application of attribute evaluation methods were iteratively added to the random forest classification beginning with the top ranking six features (age, having arthritis, having a family doctor, having diabetes, whether you get a regular checkup, having heart disease) and incrementally incorporating the next best ranking features. The classification accuracy generally improved (except for J48 decision tree) with the incorporation of each new set of features indicating that all 47 features should be considered for the final model. Table IV depicts the results when considering the top nine selected features given these were each ranked as the most impactful variables for the model as determined by the four attribute selection algorithms.

TABLE V
COMPARISON OF PERFORMANCE METRICS OVER FIVE CLASSIFIERS USING ALL 47 FEATURES. [5]

|          | Naive Bayes | J48 Decision Tree | Random Forest | Multilayer Perceptron |
|----------|-------------|-------------------|---------------|-----------------------|
| PPV      | 0.738       | 0.761             | 0.825         | 0.743                 |
| NPV      | 0.616       | 0.584             | 0.635         | 0.656                 |
| Accuracy | 0.769       | 0.623             | 0.724         | 0.789                 |
| ROC      | 0.729       | 0.701             | 0.733         | 0.768                 |

[4]Model trained on the 2011 data and tested on the 2012 data. Features considered: having heart disease, being a daily smoker, having a family doctor, having arthritis, having diabetes, having cancer, age, household income, and having been either divorced, separated or widowed.

[5]Model trained on the 2011 data and tested on the 2012 data. K* was left out due to insufficient computational resources to build and evaluate the model.

Establishing that all features should be considered for the generation of the decision trees in the random forest, we then optimized our performance metrics by varying the number of decision trees in our forest and the maximal depth of each tree. Trading off the improvement of our performance metrics with the computational complexity (determined to be the computation time for random forest generation and time required for prediction) of the model. We determined that a random forest of 25 decision trees, each limited to a depth of 20 decision node levels to be optimal. Despite these improvements, the NPV was still generally lower than the PPV, however the negative class is associated to those individuals we are most interested in, the individuals who have not recently been vaccinated. Investigating those individuals determined to be FN, we found the majority to belong to younger age groups (generally below the age of 50). Since those features with the greatest discriminability are age related (*e.g.* having heart disease, having arthritis) we hypothesized that segregating the data and training an "expert" random forest classifier for respective age groups would improve our performance metrics further. Additionally, the incorporation of additional training data from additional years, 2009 - 2013, and a new test set, 2014, was used throughout these age-segregation experiments, additionally improving our performance.

The data sets were segregated at various age-split boundaries, wherein individuals strictly above the age-split boundary were grouped together as the "older" population while the remainder belonging to or below the boundary were grouped as the "younger" population. Of the younger population, we chose to solely optimize the PPV. This was motivated by the fact that flu vaccination would be promoted for all younger individuals except those who are identified as having been recently vaccinated (the positives in our classification). This would promote the establishment and maintenance of herd immunity in the healthier age groups. Of the senior population, we chose to solely optimize the NPV. This was motivated by the fact that this would promote flu vaccination in those individuals our model identified as not having been vaccinated, and therefore (given their age) would be susceptible to further complications should they become infected. Figure 2 identifies the performance metrics of interest for each age-segregated population based on the operating split boundary.

A point of intersection appears to exist between the age-split of 60 and 70 years old. Operating with an age-split of 60 years old optimised both PPV and NPV while still maintaining approximately balanced dataset sizes. The PPV in the younger data set, when trained on all individuals under the age of 60 for the years 2009-2013 and tested on all individuals under that age in the 2014 data set was found to be 95.8%. The NPV in the older data set when trained on data containing only individuals above the age of 60 from the years 2009-2013 and tested on those above the age of 60 in the 2014 data was 77.8%. To verify that the subset of data used in training was beneficial, we compared these results to models trained on the complete, un-segregated data set, but tested on the age-split subsets. We found the NPV dropped for the older population. OnContrarily, the PPV improved slightly to 96.0%, indicating that the model can better discriminate the negative class and therefore better identify a TP. It may also be that the increase in training data improved the classification model performance.

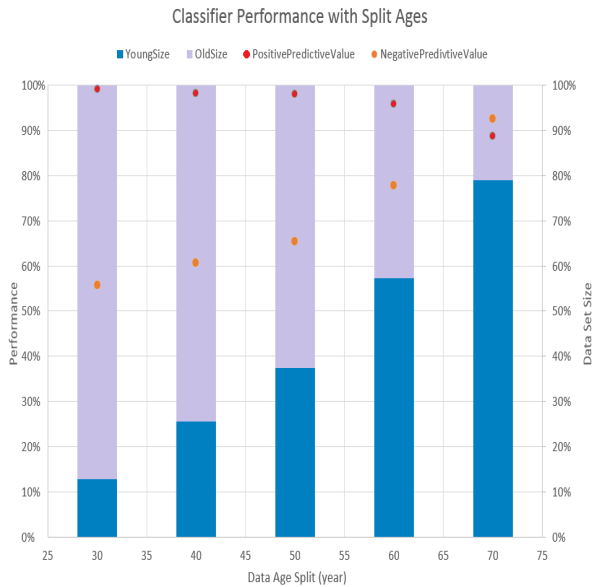

Fig. 2. PPV of the classification of younger individuals and NPV of senior individuals when training and testing on data split at varied ages.

## V. Discussion

This indicated a high similarity in the features of elder populations the discriminability when incorporating younger data into the model, which differ significantly.

Our work has used the CCHS data and identified variables related to several characteristics with the ability to predict whether or not an individual has recently been vaccinated. This work developed a model that is generalizable to the Canadian population and functions as a deployable tool to efficiently allocate vaccination promotion efforts.

### A. Interpretation of Probit Model Results

The probit model we developed identified 47 variables that were important in predicting whether someone received their flu shot. This focused on demographic variables since an individual's gender, socioeconomic status, and location (among other factors) were associated with different vaccination outcomes. Behavioural characteristics were also considered since there is an element of risk associated with failing to get a flu shot and other risk-taking behaviours (*e.g.* whether someone uses a cell phone while driving) may capture some of the risk-taking characteristics individuals in the survey had. Multiple probit models were estimated, with each iteration excluding groups of variables that were not intrinsically or statistically significant in predicting whether someone received their flu shot. The final probit model showed that factors such as province of residence, marital status, and behavioural characteristics such as whether someone has strong social networks, or has a family doctor were intrinsically and statistically significant in predicting whether someone received their flu shot. Specifically, individuals who had a family doctor were 7.35% more likely to have received a flu shot than similar individuals who did not have a family doctor. Individuals who make healthy food choices were 6.56% more likely to receive their flu shot than individuals who did not consider the health content of their food. University graduates were 2.35% more likely to receive their flu shot than individuals who did not complete high school. These marginal effects on flu vaccination outcomes are all statistically significant at the 1% significance level. These results suggest that individuals who were more educated or were less likely to engage in risk activities were more likely to receive their flu shot.

### B. Interpretation of Random Forest Model Results

The random forest classification method was robust and stable across all years considered in this study. The consistency in datasets enabled pooling of individual data for training purposes which improved our classification performance overall. The final model incorporates 25 decision trees, each limited to 20 decision nodes, randomly selecting subsets of features across all 47 identified in the probit model. In varying the number of trees, we found the performance metrics to converge when applying 25 decision trees. Very little gain in performance ($<0.1\%$) was observed, however computational time was significantly increased. Additionally, varying the limited depth of the decision trees impacted our performance. When the depth was limited ($<10$ decision nodes) our model performed poorly, indicating that the decision trees could not capture the "complexity" of the data representing an individual. This is further supported by the fact that incorporating additional decision trees to the forest had little or no impact on the classification. Similarly to the determination of decision tree number, our performance metrics converged when tree depth approached 20 levels of decision depth and was therefore set to this depth since the gain was marginal ($<0.1\%$). In order to capture the inherent complexity of this discrimination problem, our decision trees are sufficiently grown. Having established that randomly selecting our subset of features from the set of all 47 variables, our final model can robustly and stably classify our individuals.

### C. Policy Implications

The final model, depicted in Figure 3, utilizes the two expert random forest classifiers developed for both the younger and older datasets. Our model initially applies a decision node based on the individual's age. This node then determines which model to apply for classification. This model has inspired two policies for implementation and deployment of our model for the Canadian population. Each policy dictates the public health action that could be applied to each individual classified by our model.

**Policy 1:** *For high risk individuals (60+), this model can be used to target specific persons who likely have not received their flu shot.* Resources allocated to promoting flu vaccination should then be targeted to this individual on the basis that such promotion would directly benefit the individual, and indirectly benefit their surrounding community.

**Policy 2:** *For all remaining individuals, this model targets persons who are likely to get their flu shot, therefore herd immunity can be established by promoting flu vaccination in non-targets.* Given that Policy 2 applies to the younger and healthier portion of the population, the vaccination promotion efforts should be directed to the majority of individuals in communities with compromised herd immunity. In order to maximize the utility of these resources, our model can identify those individuals who have likely already been vaccinated, thereby directing vaccination promotion efforts to all individuals in the broader community that have not been identified as positives in our model.

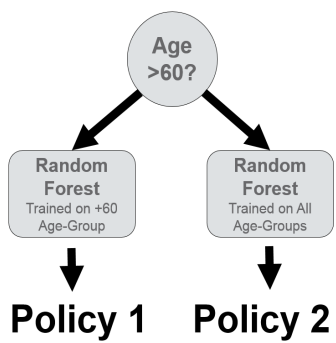

Fig. 3. Final Model. A decision node segregates individuals based on age, determining which random forest classification model to employ and the subsequent actionable decision to employ based on the outcome.

The Canadian Government budget has recently allocated $25 million dollars to promoting vaccination rates and using data sampled from communities across Canada. This methodology could identify regions with lower vaccination rates and thereby identifying them as priority regions for vaccination promotion efforts. To promote herd immunity, this model could additionally pinpoint specific individuals that have recently been vaccinated in younger populations. For individuals in elder populations at higher risk of adverse health conditions if the model could identify specific individuals that they have not been vaccinated. Using these two deployment strategies for older and younger populations, this model provides the opportunity to deploy public health resources in an efficient way to improve flu vaccination rates throughout Canada.

## VI. CONCLUSIONS

This work has used the CCHS data from the years 2009-2014 to build and test a predictive model which may be broadly applicable to the Canadian population to support flu vaccination rates. From the 1,381 available variables, 47 were identified, through a probit model, to be both statistically and intrinsically significant for the prediction of recent flu vaccination outcomes. The data from these features across individuals from each year were then used to develop and optimize two random forest classification models. The first model is trained with data belonging only to individuals with age >60 and optimizes the NPV, reported as 80.1% and through Policy 1, identifies those individuals who should be directly targeted for vaccination promotion efforts to prevent potential adverse health effects. The second model is trained on the entire range of ages and identifies with PPV of 96.4% those individuals of or below the age of 60 who have recently been vaccinated, and through Policy 2, indicates that vaccination promotion efforts should not be applied to this individual, and rather disseminated throughout the general population to improve herd immunity in the community. Combined, our final model achieves 87.8% overall accuracy and has utility to optimize the resources allocated by the Canadian government to increase vaccination rates and herd immunity.

## VII. ACKNOWLEDGEMENTS

This research would like to thank Dr. Olga Baysal and Dr. Boyan Bejanov for their guidance in this work and special acknowledgment to Dr. James Green for his feedback on this project and in the assistance in the presentation of this work at Carleton University's Data Day 3.0.